\documentclass[traditabstract]{aa}

\usepackage{graphicx,color}
\usepackage{amssymb,verbatim,hyperref}
\usepackage{txfonts}
\usepackage{natbib}
\usepackage{mathptmx}

\def\vb{\mathbf{v}}
\def\xb{\mathbf{x}}
\def\kb{\mathbf{k}}

\begin{document}

\title{XRISM constraints on the velocity power spectrum in the Coma cluster} 

\author{D. Eckert\inst{\ref{gva}} \and M. Markevitch\inst{\ref{gsfc}} \and J. A. ZuHone\inst{\ref{cfa}} \and M. Regamey\inst{\ref{gva}} \and I. Zhuravleva\inst{\ref{chic}} \and Y. Ichinohe\inst{\ref{riken}} \and N. Truong\inst{\ref{umary},\ref{gsfc}} \and N. Okabe\inst{\ref{hiro}} \and D. R. Wik\inst{\ref{utah}}}

\institute{
Department of Astronomy, University of Geneva, Ch. d'Ecogia 16, 1290 Versoix, Switzerland\label{gva}\\
\email{Dominique.Eckert@unige.ch}
\and
NASA / Goddard Space Flight Center, Greenbelt, MD 20771, USA\label{gsfc}
\and
Center for Astrophysics | Harvard \& Smithsonian, 60 Garden Street, Cambridge, MA 02138, USA\label{cfa}
\and
Department of Astronomy and Astrophysics, University of Chicago, Chicago, IL 60637, USA\label{chic}
\and
RIKEN Nishina Center, Saitama 351-0198, Japan\label{riken}
\and
Center for Space Sciences and Technology, University of Maryland, Baltimore County, 1000 Hilltop Circle, Baltimore, MD 21250, USA\label{umary}
\and
Physics Program, Graduate School of Advanced Science and Engineering, Hiroshima University, 1-3-1 Kagamiyama, Higashi-Hiroshima, Hiroshima 739-8526, Japan\label{hiro}
\and
Department of Physics \& Astronomy, University of Utah, 115 South 1400 East, Salt Lake City, UT 84112, USA\label{utah}
}

\abstract{The velocity field of intracluster gas in galaxy clusters contains key information on the virialization of infalling material, the dissipation of AGN energy into the surrounding medium, and the validity of the hydrostatic hypothesis. The statistical properties of the velocity field are characterized by its fluctuation power spectrum, which is usually expected to be well described by an injection scale and a turbulent cascade. The Resolve instrument on board \emph{XRISM} allows us for the first time to accurately measure Doppler shifts and line broadening in nearby clusters. Here we propose a simulation-based inference technique to retrieve the properties of the velocity power spectrum from X-ray micro-calorimeter data by generating simulations of Gaussian random fields from a parametric power spectrum model. We forward model the measured bulk velocities and velocity dispersions by including the most relevant observational effects (projection, emissivity weighting, PSF smearing). We then train a neural network to learn the mapping between the power spectrum parameters and the generated data vectors. Considering a three-parameter model describing turbulent energy injection on large scales and a power-law cascade, we found that two \emph{XRISM}/Resolve pointings are sufficient to accurately determine the turbulent Mach number and set interesting constraints on the injection scale. Applying our method to the Coma cluster data, we obtain a model that is characterized by a large injection scale that rivals the size of the cluster ($\ell_{inj}=2.2_{-1.0}^{+2.0}$ Mpc).  When this power spectrum model is integrated over the cluster scales ($0<\ell<R_{500}=1.4 $Mpc), the Mach number of the gas motions is $\mathcal{M}_{3D,500}=0.45_{-0.13}^{+0.18}$, which exceeds the value derived from the velocity dispersions only. Further observations covering a wider area are required to decrease the cosmic variance and constrain the slope of the turbulent cascade.}

\maketitle

\section{Introduction}

As the culmination of the structure formation process, galaxy clusters are sensitive cosmological probes through their dependence on the high-mass end of the halo mass function \citep[e.g.][]{allen11,pierre16,clerc2023}. Surveys of increasing size and depth such as \emph{eROSITA} \citep{ghirardini24}, \emph{Euclid} \citep{EuclidPrep19,EuclidQ125}, SPT \citep{bocquet24} or ACT \citep{hilton21} are now detecting many thousands of clusters, which can potentially lead to groundbreaking discoveries. At present times, the main bottleneck for these surveys lies in the characterization of the detected clusters, and specifically on the relation between the survey observables and the mass of the underlying halo \citep[see][for a review]{pratt19}. In this respect, observables based on the properties of the intracluster medium (ICM) such as temperature, gas mass, or Compton-y parameter are of particular interest, as they are expected to be tightly correlated with the halo mass \citep[e.g.][]{pratt09,ettori15,lovisari20,Sereno2020}. The primary unknown is the impact of structure formation processes and baryonic physics on these observables \citep{mccarthy10,lebrun14,Truong2018,Pop2022,Braspenning2024}, currently preventing observables to be straightforwardly derived from first principles. 

In this respect, measurements of the gas velocity field in galaxy clusters hold a key role. Indeed, numerical simulations have long predicted that a fraction of the energy in the ICM is not thermalized \citep{rasia06,lau09,nelson14,biffi16,ayromlou24}. Given the low density of the medium, the mean free path in the ICM is expected to be on the order of a kpc, although small-scale instabilities may reduce the mean free path substantially \citep{Schekochihin:2022}. As such, the kinetic energy injected by successive merging events takes several Gyr to thermalize \citep[see][for a review]{simionescu19}. On top of that, outflows from central active galactic nuclei introduce perturbations in the neighboring medium by generating shocks, carving cavities, and driving turbulence, which affects the thermodynamic properties of the ICM and its gas distribution \citep{gaspari11,gaspari15,bourne17,wittor20,bourne21}. Numerical simulations predict that most of the remaining kinetic energy should be in the form of random gas motions \citep{lau09,zuhone13b,va17turb,angelinelli20,pearce20,barnes21,bennett22,groth25}. Perturbations injected on large scales generate a turbulent cascade and dissipate on small scales where thermalization becomes effective \citep[e.g.][]{gaspari14,zhuravleva14}. The statistical properties of the velocity field can be characterized by a fluctuation power spectrum that peaks on large scales at the so-called injection scale and decreases on smaller scales as a power law with an index close to the classical Kolmogorov slope of 11/3 \citep{schuecker,churazov12,gaspari13,va17turb,zhuravleva18,simonte22}. 

Observational constraints on the velocity fluctuation power spectrum are currently scarce due to the lack of instruments with appropriate capabilities. Studies of line profiles in \emph{XMM-Newton}/RGS observations of cool-core clusters only yielded mild constraints on the turbulence of the ICM \citep{sanders11,sanders13,pinto15,ogorzalek17}. More recently, \citet{sanders20} proposed a new technique to accurately calibrate the energy gain of the \emph{XMM-Newton}/EPIC-pn CCD, which resulted in measurements of bulk velocities in a handful of nearby clusters \citep{gatuzz22,gatuzz23,gatuzz24}, although the accuracy of the method is limited to $\sim100$ km/s. Alternatively, several works used the fluctuation field of thermodynamic quantities (density, pressure, temperature) as an indirect tracer of the velocity field \citep{schuecker,churazov12,gaspari14,zhuravleva14,hofmann16,khatri16}. These studies imply that the turbulent pressure support in galaxy clusters is relatively mild \citep[$P_{\rm turb}/P_{\rm tot}\lesssim0.2$,][]{zhu15,eckert17,dupourque23,dupourque24,romero23,romero24,heinrich24,maringilabert24,lovisari24}. However, the relation between the thermodynamic fluctuations and the underlying velocity field is uncertain, most notably in merging clusters \citep{simonte22}, in large part due to contamination by other unrelated processes such as subhalos, gas sloshing, or ellipticity \citep{zhuravleva23}.

The field is expected to advance significantly thanks to the advent of X-ray micro-calorimeters. The first observation of a galaxy cluster with a micro-calorimeter was obtained in 2016 by the SXS instrument on board \emph{Hitomi}, which targeted the Perseus cluster as its first-light observation \citep{hitomi16,hitomi18}. In the cluster core, the Fe XXV emission line complex was found to be broadened by 160 km/s, which, if interpreted as evidence for random isotropic motions, amounts to a small fraction ($\sim4\%$) of the thermal pressure \citep{hitomi16}. While \emph{Hitomi} was unfortunately lost shortly thereafter, its successor \emph{XRISM} \citep{tashiro25} was launched in September 2023 and has been successfully operating since then. \emph{XRISM} carries the Resolve instrument, an array of $6\times6$ individual pixels featuring a spectral resolution of $4.5$ eV. \emph{XRISM} observed several clusters thus far \citep{xrismcoma,xrismcentaurus,xrisma2029}, revealing moderate line widths (100-200 km/s) but substantial bulk motions as large as 700 km/s in Coma \citep[hereafter X25]{xrismcoma}. However, linking these measurements to the underlying velocity power spectrum is far from trivial, which makes the comparison with predictions from numerical simulations difficult.

In this paper, we propose a simulation-based inference (SBI) method to determine the power spectrum of velocity fluctuations in galaxy clusters based on X-ray micro-calorimeter data. We generate simulations of the velocity field for any given power spectrum shape, and fold in observational effects (emissivity weighting, projection, PSF smearing, measurement uncertainties) to generate mock bulk velocities and velocity dispersions for any observational setup. We then train a neural network to learn the relation between the parameters describing the 3D power spectrum and the mock datasets. The neural network can then be used in reverse to infer the model parameters given the observed data. We validate our method using extensive simulations and apply it to the measured bulk velocities and velocity dispersions in the Coma cluster. The paper is organized as follows. In Sect. \ref{sec:sim_pipe} we describe the simulation pipeline and the optimization procedure. We validate our method in Sect. \ref{sec:recovery} and apply it to the \emph{XRISM}/Resolve data of the Coma cluster in Sect. \ref{sec:coma}, comparing the results with the constraints obtained from the same data in X25. We discuss our results in Sect. \ref{sec:disc}.

Throughout the paper, we assume a flat $\Lambda$CDM cosmology with $\Omega_m=0.3$ and $H_0=70$ km/s/Mpc. At the mean redshift of the Coma cluster \citep[$z=0.0233$,][]{bilton18}, this corresponds to $1^\prime=28.5$ kpc.

\section{Methodology}
\label{sec:meth}

In this section we present the theoretical framework that we use to forward model the velocity measurements and our SBI implementation. We then validate the technique using a large set of simulations and study the ability of our method to recover the parameters of the underlying velocity power spectrum. We distribute our code in the form of an easy-to-use Python package\footnote{\href{https://gitlab.astro.unige.ch/eckertd/clusterps/}{https://gitlab.astro.unige.ch/eckertd/clusterps/}}.

\subsection{Simulation setup}
\label{sec:sim_pipe}

\begin{figure*}
\resizebox{\hsize}{!}{\includegraphics[width=0.54\textwidth]{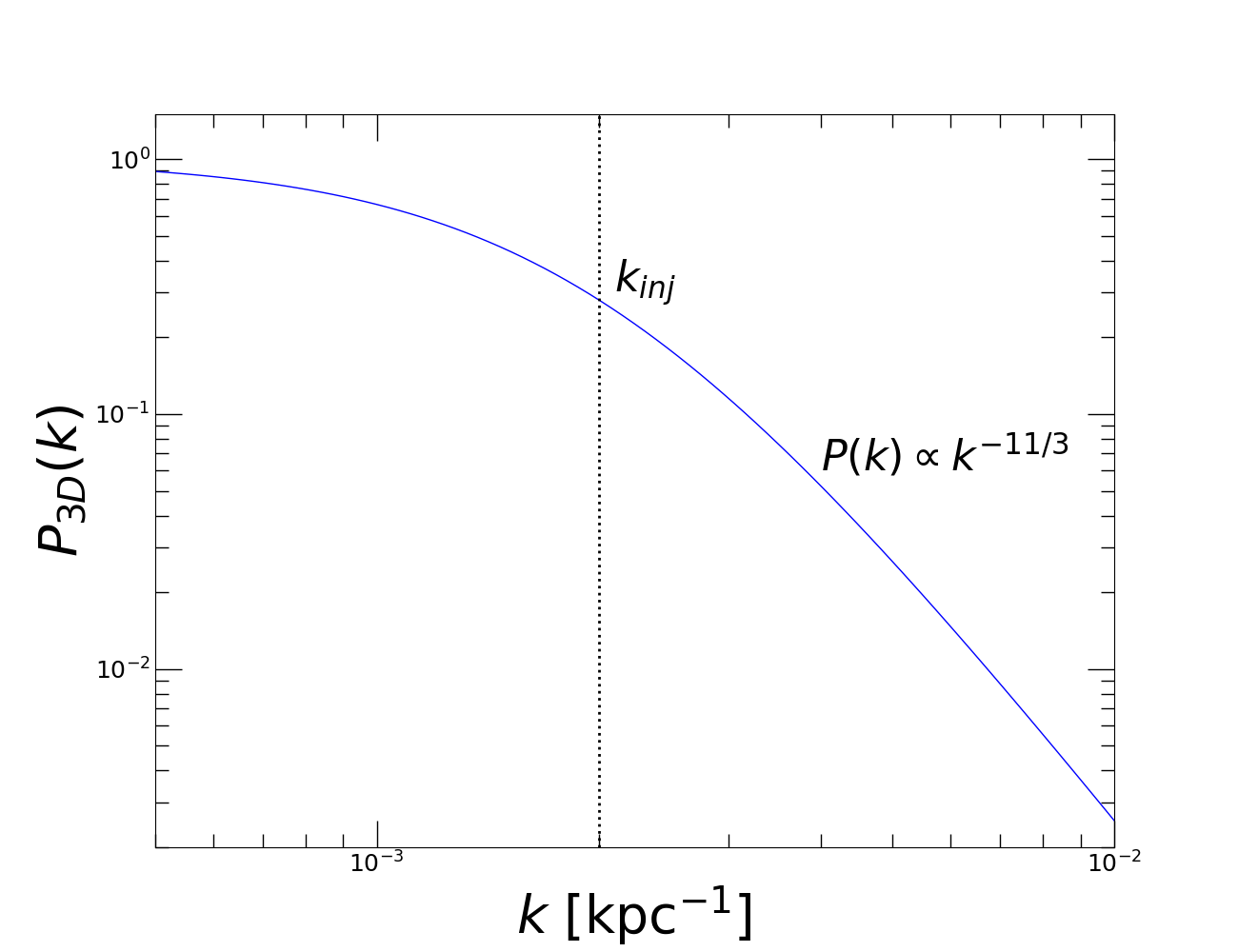} \includegraphics[width=0.46\textwidth]{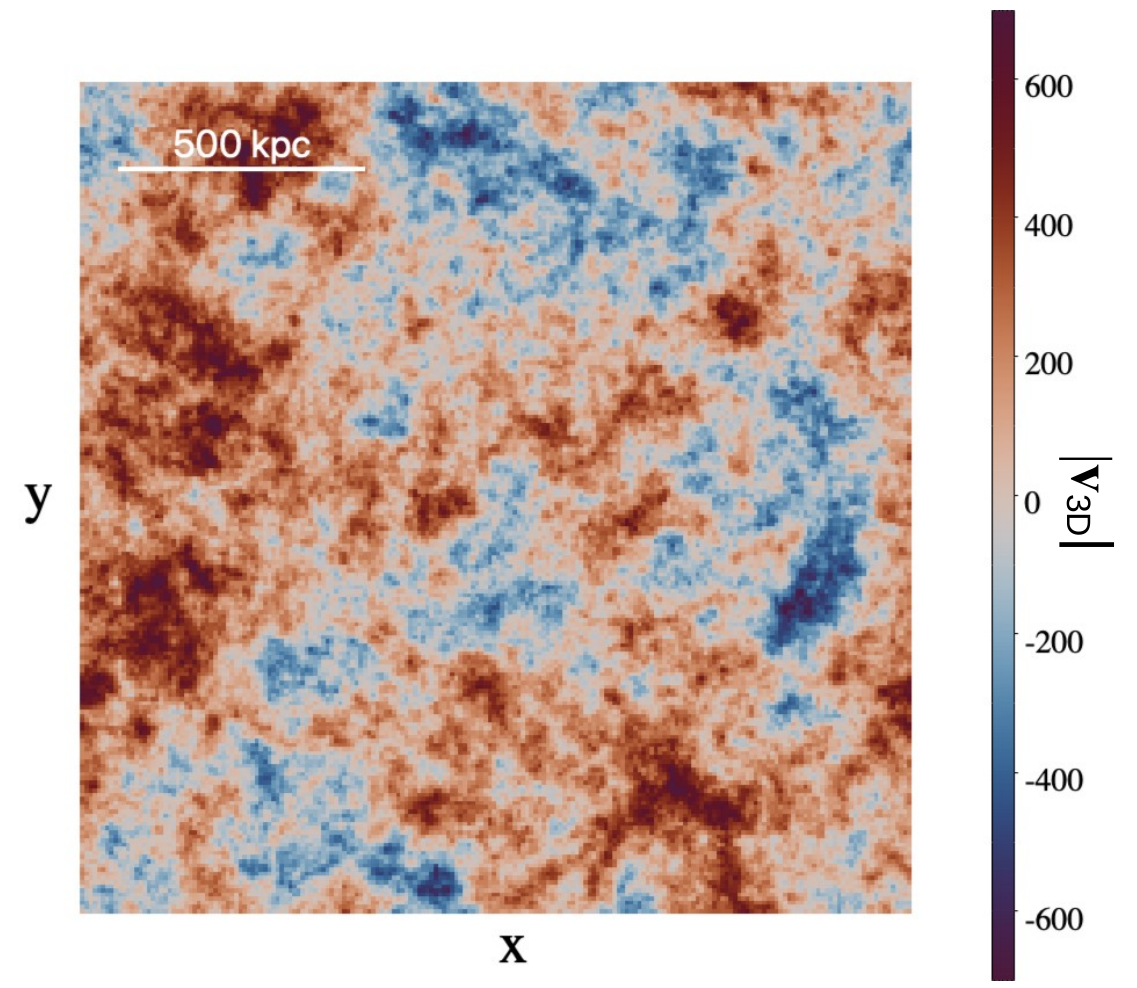}}
\caption{\label{fig:sketch} Example of fluctuation field generation. \emph{Left:} Model 3D power spectrum described by an injection at $k_{inj}$ and a classical Kolmogorov cascade on smaller scales (Eq. \ref{eq:classicalps}). The vertical dashed line shows the position of $k_{inj}=({\rm 500 kpc})^{-1}$. \emph{Right:} Slice of the real-space velocity fluctuation field along the $z$ axis generated from this power spectrum (Eq. \ref{eq:realspacevel}) for a velocity dispersion $\sigma_v=300$ km/s.  The color code shows the velocity in km/s. The box size is 1.7 Mpc on a side.}
\end{figure*}

We consider a three-dimensional velocity field $\vb(\xb) = (v_x(x,y,z), v_y(x,y,z), v_z(x,y,z))$, with $x,y$ the Cartesian coordinates on the plane of the sky and $z$ the coordinate along the line of sight. The three-dimensional power spectrum of the velocity field is defined as the square modulus of the velocity field in Fourier space,
\begin{equation}
P_{3D}(\kb) = \left| \int \vb(\xb) e^{-i2\pi \kb\cdot\xb} ~ {\rm d}^3x \right|^2,
\end{equation}
\noindent with $\kb=(k_x,k_y,k_z)$ the wave number. Assuming isotropy, the power spectrum can be reduced to the one-dimensional function $P_{3D}(k)$, with $k=|\kb|=1/\ell$. 

For a simulation box of size $N^3$, we set linearly spaced wave numbers with the origin set at the center of the box, such that each axis ($k_x,k_y,k_z$) is linearly spaced and spans values in the range $k \in (-N/2, N/2)$. Given a parametric model for $P_{3D}$, we calculate the power $P_{3D}(\kb)$ for every point on the grid and generate a random Gaussian realization of the Fourier transform $\hat \vb$ of the velocity field as 
\begin{equation}
\hat \vb(\kb) = P_{3D}(\kb)^{1/2} \mathcal{N}_C,
\end{equation}

\noindent with $\mathcal{N}_C$ a random complex number with normally distributed real and imaginary parts. The real-space velocity field can then be obtained as the real part of the inverse Fourier transform of $\hat \vb$,

\begin{equation}
\vb(\xb) = \operatorname{Re} \left[ \int \hat\vb(\kb) e^{i2\pi\kb\cdot\xb} ~ {\rm d}^3k\right].
\label{eq:realspacevel}
\end{equation}

This procedure generates a 3D velocity field centered on zero with Gaussian fluctuations representing isotropic turbulent eddies. As an example, in Fig. \ref{fig:sketch} we show a realization of the velocity field from a power spectrum given by the classical expression for a turbulent cascade generated by injection on large scales \citep[e.g.][]{zhuravleva12,zuhone16},
\begin{equation}
P_{3D}(k) = \sigma_v^2 \frac{(1+(k/k_{inj})^2)^{-\alpha/2}}{\int 4\pi k^2 (1+(k/k_{inj})^2)^{-\alpha/2} ~{\rm d}k}.
\label{eq:classicalps}
\end{equation}

Here $k_{inj}$ describes the wave number corresponding to the energy injection scale, $\alpha$ is the slope of the turbulent cascade, and the model is normalized such that the amplitude of the spectrum is equal to the variance $\sigma_v^2$ of the velocity field. The left-hand panel of Fig. \ref{fig:sketch} shows the power spectrum described by Eq. \ref{eq:classicalps} for an injection scale $\ell_{inj}=(k_{inj})^{-1}=500$ kpc, a slope of $11/3$ corresponding to the classical Kolmogorov slope, and an arbitrary normalization. In the right-hand panel we show a slice of the 3D velocity field along the $z$ axis for a realization of a Gaussian random field characterized by this power spectrum, with a velocity dispersion $\sigma_v=300$ km/s, a pixel size of 8 kpc, and a box size of 1.7 Mpc on a side. We can see that the procedure generates fluctuations on all scales with a maximum around $\ell_{inj}=500$ kpc, as expected from the model of choice.

Given a paramterization for $P_{3D}(k)$ and a set of model parameters, we generate a real-space fluctuation field $\vb(\xb)$ according to the procedure described above. Since measurements of Doppler shifts and velocity dispersions are obtained along the line of sight ($z$ axis), we need to project the velocity field to predict the measured quantities. Let $EM_{3D}(\xb)$ be the three-dimensional emissivity distribution, which is proportional to the squared gas density. The projected bulk velocity is given by the mean of the 3D velocity along the $z$ direction, weighted by the local gas emissivity,
\begin{equation}
v_{b}(x,y) = \frac{\int EM_{3D}(\xb)v_z(\xb)~{\rm d}z}{\int EM_{3D}(\xb)~{\rm d}z}.
\label{eq:vbulk1D}
\end{equation}
Similarly, the line-of-sight averaged velocity dispersion $\sigma(x,y)$ can be estimated as the quadratic mean of the velocity fluctuations along the $z$ axis, weighted by the local emission measure,
\begin{equation}
\sigma(x,y) = \left[\frac{\int EM_{3D}(\xb)(v_z(\xb)-v_{b}(x,y))^2~{\rm d}z}{\int EM_{3D}(\xb)~{\rm d}z} \right]^{1/2}
\label{eq:sigmav1D}
\end{equation}
\citet{truong24} demonstrated with mock spectra from simulated clusters from the TNG-Cluster suite that Eqs. \ref{eq:vbulk1D} and \ref{eq:sigmav1D} provide a good description of the quantities estimated from X-ray spectroscopy \citep[see also][]{roncarelli18}. 

To describe the emission measure distribution $EM_{3D}(\xb)$, we assume that the system of interest is spherically symmetric with a gas distribution described by a spherical $\beta-$model \citep{cavaliere}, 
\begin{equation}
EM_{3D}(r) = EM_0 \left(1 + \left(\frac{r}{r_c}\right)^2 \right)^{-3\beta}.
\label{eq:betamodel}
\end{equation}
This model provides a good representation of the surface brightness profile of the Coma cluster \citep[e.g.][]{churazov12}. Fitting Eq. \ref{eq:betamodel} to the \emph{XMM-Newton} surface brightness profile of the system yields $r_c=266$ kpc and $\beta=0.66$. The choice of the $\beta-$model is motivated by the fact that the relation between 2D and 3D distributions is analytic \citep[e.g.][]{eckert20}. However, any model of choice can be substituted in case the $\beta-$model is not sufficiently accurate. 

In the presence of a non-negligible instrumental point spread function (PSF), as is the case for \emph{XRISM}/Resolve, the measured bulk velocities and velocity dispersions need to be averaged out also along the plane-of-the-sky coordinates ($x,y$), again weighted by the local emission measure. Equation \ref{eq:vbulk1D} then becomes
\begin{equation}
v_{b}(x^\prime,y^\prime)=\frac{\int PSF(x-x^\prime,y-y^\prime) EM_{3D}(\xb) v_z(\xb)~{\rm d^3}x}{\int PSF(x-x^\prime,y-y^\prime) EM_{3D}(\xb)~{\rm d^3}x}.
\label{eq:psf}
\end{equation}
\noindent Eq. \ref{eq:sigmav1D} can be modified in a similar way to predict the velocity dispersion in the presence of a relatively broad instrumental PSF. In the case of \emph{XRISM}/Resolve, we model the PSF as a two-dimensional Gaussian with $\sigma=34^{\prime\prime}$, which gives a reasonable description of the telescope's PSF \citep{tashiro25}. However, any PSF shape can be considered and folded into Eq. \ref{eq:psf}. 

Finally, the observational configuration can be introduced, i.e. the regions for which bulk velocity and velocity dispersion measurements were obtained can be defined as a function of the plane-of-the-sky coordinates $(x,y)$. The measurements of bulk velocities and velocity dispersions within each region of total area $\Omega$ are then obtained by integrating Eq. \ref{eq:psf} over the defined area, i.e.
\begin{equation}
v_{1D} = \frac{1}{\Omega} \int v_b(x,y)~{\rm d}\Omega.
\end{equation}
The final values of $v_{1D}$ and $\sigma_{1D}$ are drawn from Gaussian distributions according to the measurement uncertainties in each region to generate a mock observational dataset which mimics the real procedure as closely as possible.

\subsection{Optimization}
\label{sec:optim}

\begin{figure*}
\resizebox{\hsize}{!}{\includegraphics{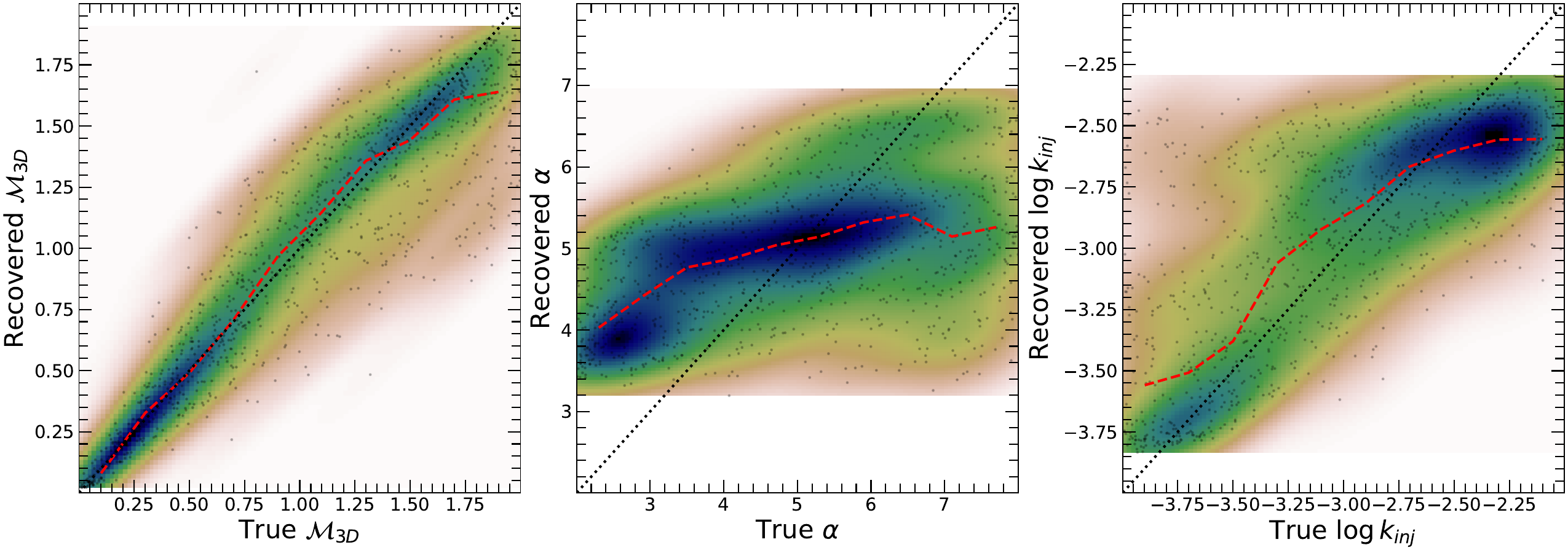}}
\caption{\label{fig:coverage} Coverage tests for the two-pointing configuration and the model defined in Eq. \ref{eq:classicalps}. All the panels show the median of the inferred posterior distribution as a function of the true input value. The panels show the recovery tests for the Mach number $\mathcal{M}_{3D}=\sigma_v/c_s$ (\emph{left}), the slope of the turbulent cascade $\alpha$ (\emph{middle}), and the logarithm of the injection wave number $k_{inj}$, in units of kpc$^{-1}$ (\emph{right}). The gray dots show the results of individual simulations, whereas the color code indicates the point density inferred using a Gaussian kernel density estimator. The dotted black lines and the red dashed lines indicate the $1:1$ relation and the running median of the data points, respectively.}
\end{figure*}

As our test configuration, we consider the two \emph{XRISM}/Resolve pointings of the Coma cluster presented in X25. The dataset consists in a deep (400ks) pointing located close to the X-ray emission peak and a second slightly shallower pointing (160 ks) located 170 kpc South of the cluster center. Each pointing was split into four approximately independent sub-arrays of 1.5 arcmin (42 kpc) on a side, such that the considered dataset includes 8 individual bulk velocity measurements with an uncertainty of 30 km/s for the central pointing and 60 km/s for the South pointing. The bulk velocities are set with respect to the cluster rest frame defined as the mean redshift of $\sim600$ member galaxies. We consider the velocity dispersions from the four sub-arrays for the deep central pointing, while for the shallower South pointing we only include a single velocity dispersion measurement for the entire array. As a result, our simulation pipeline generates a data vector that contains a set of 8 bulk velocities and 5 velocity dispersions. 

The procedure described in Sect. \ref{sec:sim_pipe} is inherently random, as an infinity of real-space fluctuation fields can be described by the same power spectrum. In particular, for two realizations generated from the same parameters, the local bulk velocities vary substantially, as the distribution of local bulk velocities is randomly distributed around zero. This effect is known as cosmic variance \citep{clerc19}. For this reason, the data cannot be compared to the model using an analytic likelihood. To alleviate this issue, we adopt simulation-based inference as our fitting technique, as was done by \citet{dupourque23,dupourque24} in the context of surface brightness fluctuations. Specifically, we use the \texttt{sbi} Python package \citep{sbi} and the sequential neural posterior estimation (SNPE) method to fit the data. For a given parameterization of $P_{3D}(k)$ (e.g. Eq. \ref{eq:classicalps}), we set a uniform prior range for the model parameters and sample parameter sets from the prior distribution. We then generate mock datasets using the simulation pipeline outlined in Sect. \ref{sec:sim_pipe} and train a neural network to describe the relation between the model parameters and the multi-dimensional posterior density distribution. We use the automatic posterior transformation method \citep{greenberg19} and the masked autoregressive flow (MAF) normalizing flow method \citep{papamakarios17} for density estimation. We refer to these papers for the details on the architecture of the network and the training method. We infer the posterior distribution of the model parameters by sampling the trained density distribution given the observed data vector. For more details on the optimization technique, we refer the reader to \citet{regamey25}.

Alternatively, following \citet{molin25} we consider a summary statistic for the bulk velocity field in the form of the velocity structure function \citep[VSF, e.g.][]{zuhone16,roncarelli18,gatuzz23b}. The VSF is defined as the mean squared difference of all velocity measurements separated by a distance $r$,
\begin{equation}
VSF(r) = \left\langle\left(v_{b}(x,y) - v_{b}(x+r_x,y+r_y)\right)^2\right\rangle,
\label{eq:vsf}
\end{equation}
\noindent with $(r_x,r_y)$ any 2D vector of modulus $r$. The VSF can be analytically related to the power spectrum \citep{zhuravleva12,zuhone16,clerc19}, such that a likelihood function can be constructed and optimized, as was done in X25 for the Coma cluster. The main drawback is that the VSF only depends on relative bulk velocities rather than absolute ones, thus it is insensitive to bulk motions occurring on scales larger than the covered observational footprint. In the following we consider both the case where we directly forward fit the bulk velocities and the case where we fit a summary statistic in the form of the VSF. In the latter case, for each simulated mock dataset, we calculate the VSF using Eq. \ref{eq:vsf}. The fitted data vector is then made of the combination of the VSF points and the velocity dispersion measurements.

\subsection{Recovery tests}
\label{sec:recovery}

We ran recovery tests with random points to test the accuracy of our method and its ability to recover the true underlying power spectrum. We considered the two-pointing configuration described in Sect. \ref{sec:optim} and an emission measure profile that is well matched to the emissivity distribution in the Coma cluster (Eq. \ref{eq:betamodel}, see Appendix \ref{sec:sb}). We considered boxes with $N=200$ pixels on a side covering a range of $1.7$ Mpc, which is much larger than the projected distance between the two pointings (170 kpc) and the effective depth along the line of sight ($\sim600$ kpc). This configuration also provides a spatial resolution of 8 kpc, which substantially oversamples the instrumental PSF ($\sigma=34^{\prime\prime}\approx16$ kpc). We assumed that the power spectrum can be described by the three-parameter model shown in Eq. \ref{eq:classicalps} and Fig. \ref{fig:sketch}. The model does not include dissipation on small scales, as the resolution of \emph{XRISM}/Resolve ($1.5^\prime\sim43$ kpc) largely exceeds the Coulomb mean free path in the medium ($\sim10$ kpc). The model parameters are the 3D velocity dispersion $\sigma_v$, the injection wave number $k_{inj}$, and the slope of the turbulent cascade $\alpha$. Instead of the velocity dispersion, we define the model as a function of the Mach number $\mathcal{M}_{3D}=\sigma_v/c_s$, with $c_s=\left(\gamma k_BT/\mu m_p\right)^{1/2}$ the sound speed in the medium. For a temperature of $8.37$ keV in the core of Coma (see X25), the sound speed amounts to $1,508$ km/s. We set uniform priors on the model parameters spanning the range $\mathcal{M}_{3D}\sim\mathcal{U}(0,2)$, $\log k_{inj} \sim\mathcal{U}(-4,-2)$, $\alpha\sim\mathcal{U}(2,8)$. These prior ranges are expected to be broad enough to encompass any reasonable configuration. 

We sampled 10,000 parameter sets from the prior distributions and generated mock data vectors for each parameter set using the simulation pipeline defined in Sect. \ref{sec:sim_pipe}. We then used the SNPE method in \texttt{sbi} to estimate the posterior density distribution. The training was achieved by maximizing a loss function using a maximum-likelihood algorithm, which converged after 148 epochs. We generated another set of 1,000 simulations and applied the model to the mock data vectors to infer the best fitting parameters. In each case, we adopted the median of the posterior distribution as our point estimate, as it is a robust estimator in the presence of skewed, non-Gaussian posteriors. We compared the resulting parameter values to the input values that were used to generate the simulations. In Fig. \ref{fig:coverage} we show the resulting coverage tests for the model defined in Eq. \ref{eq:classicalps} and the two-pointing configuration. The gray points show the median of the recovered posterior distributions for each individual simulation, whereas the smooth color scale shows the point density estimated using a Gaussian kernel density estimator. We can see that the method accurately recovers the 3D Mach number, with an uncertainty ranging from $\sim0.1$ at low Mach numbers to $\sim0.25$ for $\mathcal{M}_{3D}>1.5$. The recovered points are unbiased throughout most of the considered range, with the exception of the very high Mach number regime where the posterior distribution hits the edge of the prior. Conversely, for the slope $\alpha$ the recovered values are close to the median of the prior ($\alpha=5$), which shows that the parameter is poorly constrained. This is not surprising, as the two-pointing configuration only probes a limited range of independent scales (from $\sim2$ to $\sim7$ arcmin) and the available local measurements are affected by cosmic variance due to the limited azimuthal coverage. Additional pointings covering a wider range of scales are needed to set constraints on the power spectrum slope. The recovery test for the injection scale $k_{inj}$ shows that the two-pointing configuration is sensitive to this parameter, as the distribution of the recovered values lies close to the one-to-one relation, although the measurement is not as accurate as for the Mach number. The injection scale can be estimated with a modest precision of 0.2-0.3 dex.

We performed similar tests for the case where the fitted data points include the VSF and the velocity dispersions instead of the individual bulk velocities \citep[see][and Sect. \ref{sec:optim}]{molin25}. Fitting for the VSF yields noticeably poorer constraints. The posterior Mach number distributions are highly skewed, with a peak that is biased towards low values and a long tail towards high Mach numbers. This is due to the VSF measurement being sensitive only to relative motions (see Eq. \ref{eq:vsf}) whereas the addition of the absolute bulk velocities with respect to the cluster rest frame yields much tighter constraints on the Mach number and the injection scale. Nonetheless, the VSF remains a very useful summary statistic to compare the best fitting model to the measurements, as it is less affected by the random nature of the process than the individual measurements.

\section{Results}

\subsection{\emph{XRISM}/Resolve observations of the Coma cluster}
\label{sec:coma}

\begin{figure}
\resizebox{\hsize}{!}{\includegraphics{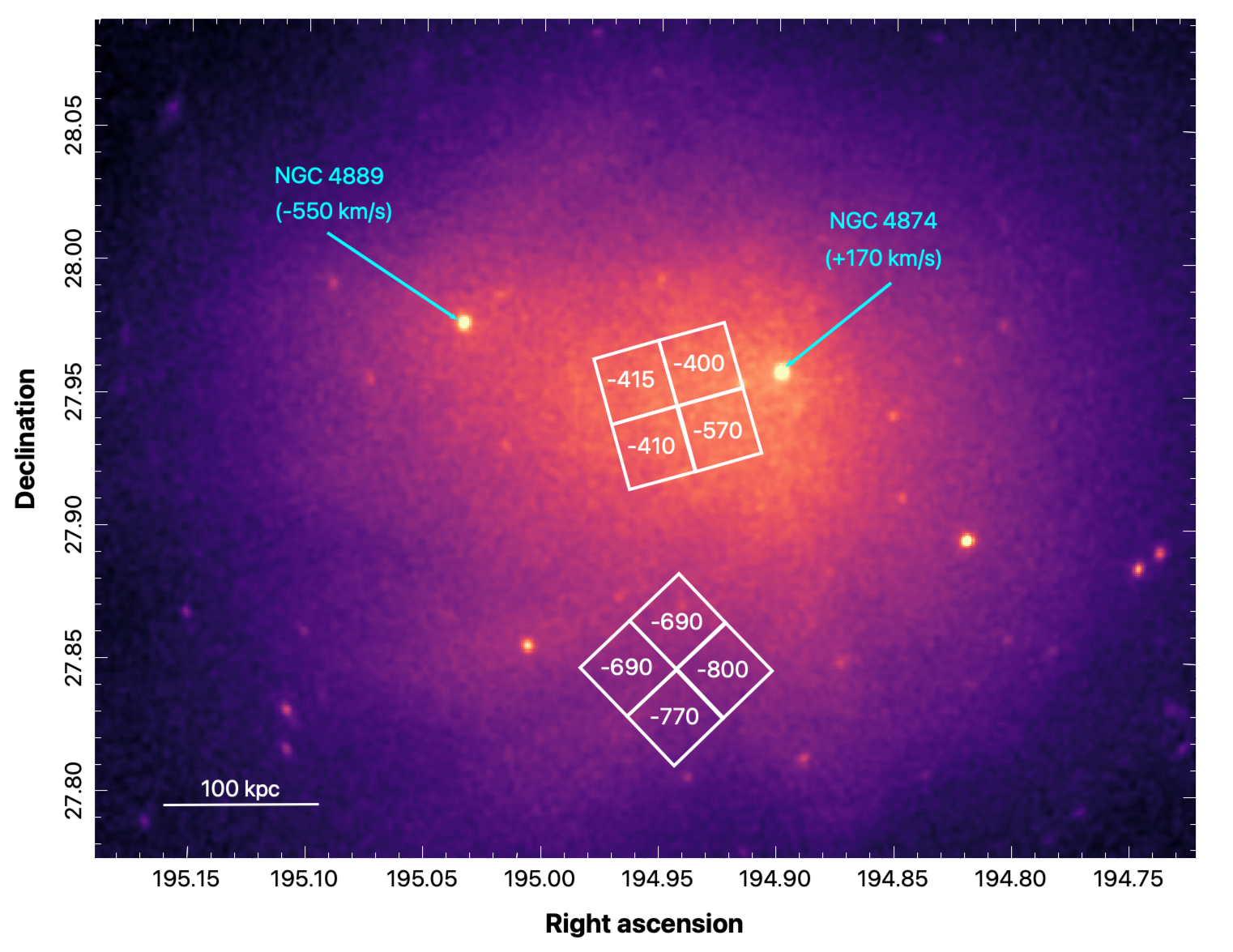}}
\caption{\label{fig:coma_map} Location of the \emph{XRISM}/Resolve measurements superimposed on a \emph{XMM-Newton}/EPIC surface brightness map of the Coma cluster. The white squares show the footprint of the two Resolve pointings split into four individual quadrants, with the number showing the bulk velocity of each region with respect to the cluster rest frame. The blue arrows show the positions of the two dominant galaxies NGC 4874 and NGC 4889 with their respective peculiar velocities. Figure reproduced with modifications from X25.}
\end{figure}

We applied the technique devised in Sect. \ref{sec:meth} to the measurements of bulk velocities and velocity dispersions obtained by \emph{XRISM}/Resolve for the Coma cluster (see X25). The position of the two \emph{XRISM} pointings is highlighted in Fig. \ref{fig:coma_map} and superimposed onto an \emph{XMM-Newton}/EPIC map of the cluster in the [0.7-1.2] keV band. Each of the two pointings was subdivided into four approximately independent quadrants, which in the end provides us with a set of eight bulk velocity measurements. The retrieved bulk velocities with respect to the cluster rest frame $z=0.0233$, set as the mean of the redshift distribution of $\sim600$ individual galaxy redshifts \citep{bilton18}, are shown in Fig. \ref{fig:coma_map}. Coma is known to contain two dominant galaxies, NGC 4874 and NGC 4889, that are separated by $\sim200$ kpc on the plane of the sky and exhibit a velocity difference of $\sim 700$ km/s. This observation has been used as evidence that the cluster is currently in a post-merger phase. Using \emph{XMM-Newton}/EPIC-pn, \citet{sanders20} found that the bulk velocities of the cluster show a bimodal behavior, with the central, southern and western regions aligning with the redshift of NGC 4889, whereas the remaining regions tend to align with NGC 4874. The Resolve measurements confirmed this picture with a much higher precision, showing that the gas in the core of the cluster exhibits considerable bulk velocities with respect to the cluster rest frame. The centroid of the lines was found to be blueshifted by 450 km/s and 730 km/s in the central and southern pointings, respectively. The gas velocities align with the redshift of NGC 4889, which identifies NGC 4889 as the dominant galaxy of the primary merging component. The velocity dispersion of the gas was found to be $210$ km/s for the central pointing and $200$ km/s for the southern region, which is substantially lower than the measured bulk velocities and possibly implies a steep velocity power spectrum. For more details on the analysis of the \emph{XRISM}/Resolve data, we refer the reader to X25. 

\subsection{Power spectrum inference}
\label{sec:inference}

\begin{figure}
\resizebox{\hsize}{!}{\includegraphics{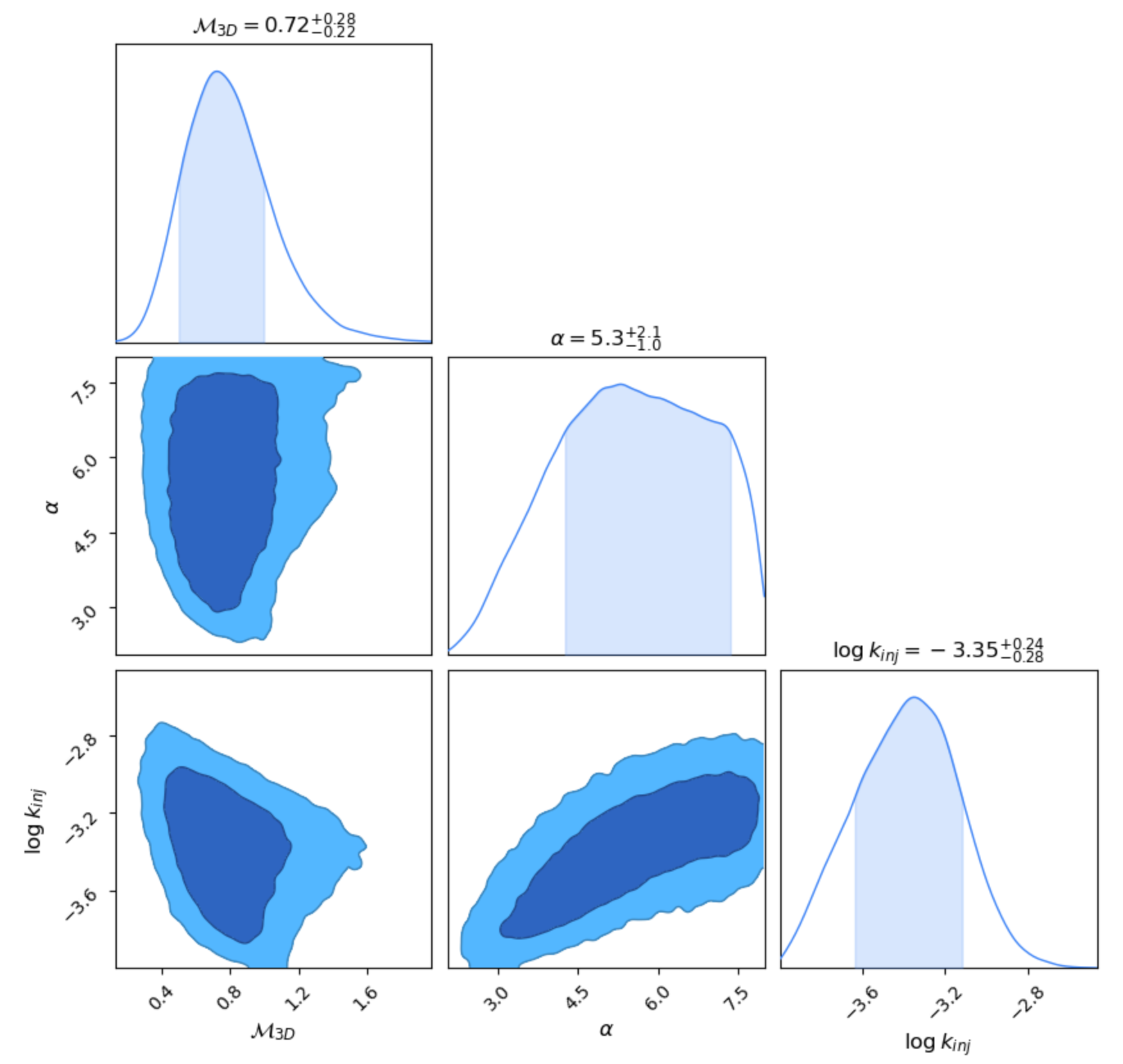}}
\caption{\label{fig:posterior} Posterior distributions for the three-parameter model (Eq. \ref{eq:classicalps}) for the Coma cluster inferred from the \emph{XRISM}/Resolve bulk velocity and velocity dispersion data. }
\end{figure}

We applied the model trained for the two-pointing configuration tuned to the Coma cluster observations and defined in Sect. \ref{sec:optim} to the data vector made of the bulk velocity and velocity dispersion measurements reported in X25. In Fig. \ref{fig:posterior} we show a corner plot extracted from the inferred probability distributions for the three-parameter model (Eq. \ref{eq:classicalps}). 

The model retrieves a fairly high Mach number $\mathcal{M}_{3D,\rm tot}=0.72_{-0.22}^{+0.28}$, which is necessary to reproduce the large bulk velocities measured in the South pointing. If these large bulk velocities are interpreted as isotropic fluctuations, the measured 1D velocities in the southern field correspond to 3D velocities of $\sim1250$ km/s, i.e $\sim80\%$ of the sound speed in the medium. The model prefers a large injection scale $\ell_{inj}=2.2_{-1.0}^{+2.0}$ Mpc. A large injection scale is required to match at the same time the large bulk velocities and the relatively low velocity dispersions. Reducing the injection scales to lower values ($<500$ kpc) increases the velocity dispersion at fixed Mach number, as the power is shifted to smaller scales. The retrieved injection scale is very large compared to the projected distance between the two pointings (170 kpc). The measurement is rendered possible by our prior knowledge of the mean cluster redshift from the galaxies and the dependence of the velocity dispersion on the line-of-sight depth of the system. 

The slope of the power spectrum is poorly constrained, as expected from the coverage test presented in Fig. \ref{fig:coverage}. As already noted in X25, the model shows a preference for a steep slope, although the classical Kolmogorov slope is well within the $1\sigma$ contours. We note that the slope and the injection wave number are positively correlated. Indeed, models featuring a smaller injection scale but a steep slope also place most of the power on large scales, which is required to generate large bulk velocities without contributing to the velocity dispersion. 

Given the weak constraints on the power spectrum slope, we also fitted the data with a simpler model whereby the slope $\alpha$ is fixed to the Kolmogorov value of 11/3, which is consistent with the slope of the density fluctuation power spectrum within the uncertainties \citep{zhuravleva19}. While the Mach number is left unchanged ($\mathcal{M}_{3D,\rm tot}=0.73_{-0.20}^{+0.25}$), the injection scale is pulled to even larger values ($\ell_{inj}=6.6_{-2.2}^{+2.5}$ Mpc) and the posterior hits the upper boundary of the prior (10 Mpc). This is expected from the correlation between $\alpha$ and $k_{inj}$ shown in Fig. \ref{fig:posterior}, where we can see that fixing $\alpha=11/3$ enforces lower values of $k_{inj}$ close to the lower edge of the prior. 

Finally, to determine the constraints that can be obtained from XRISM data only, we ran the power spectrum reconstruction by fitting the VSF instead of the individual bulk velocities as discussed in Sect. \ref{sec:optim}. In this case, we retrieve statistically consistent results, although the posterior distribution of the Mach number ($\mathcal{M}_{3D}=0.36_{-0.12}^{+0.50}$) is found to be highly skewed, with a peak at a substantially lower value and an extended tail towards high values essentially corresponding to a lower limit. This result can be explained by the lack of constraints on the absolute bulk velocities, as the VSF only depends on relative bulk velocities. The results on the injection scale and slope are consistent with the values obtained by fitting the individual bulk velocity values and independently confirm the need for a large injection scale.

\subsection{Goodness of fit}
\label{sec:goodness}

\begin{figure*}
\resizebox{\hsize}{!}{\includegraphics{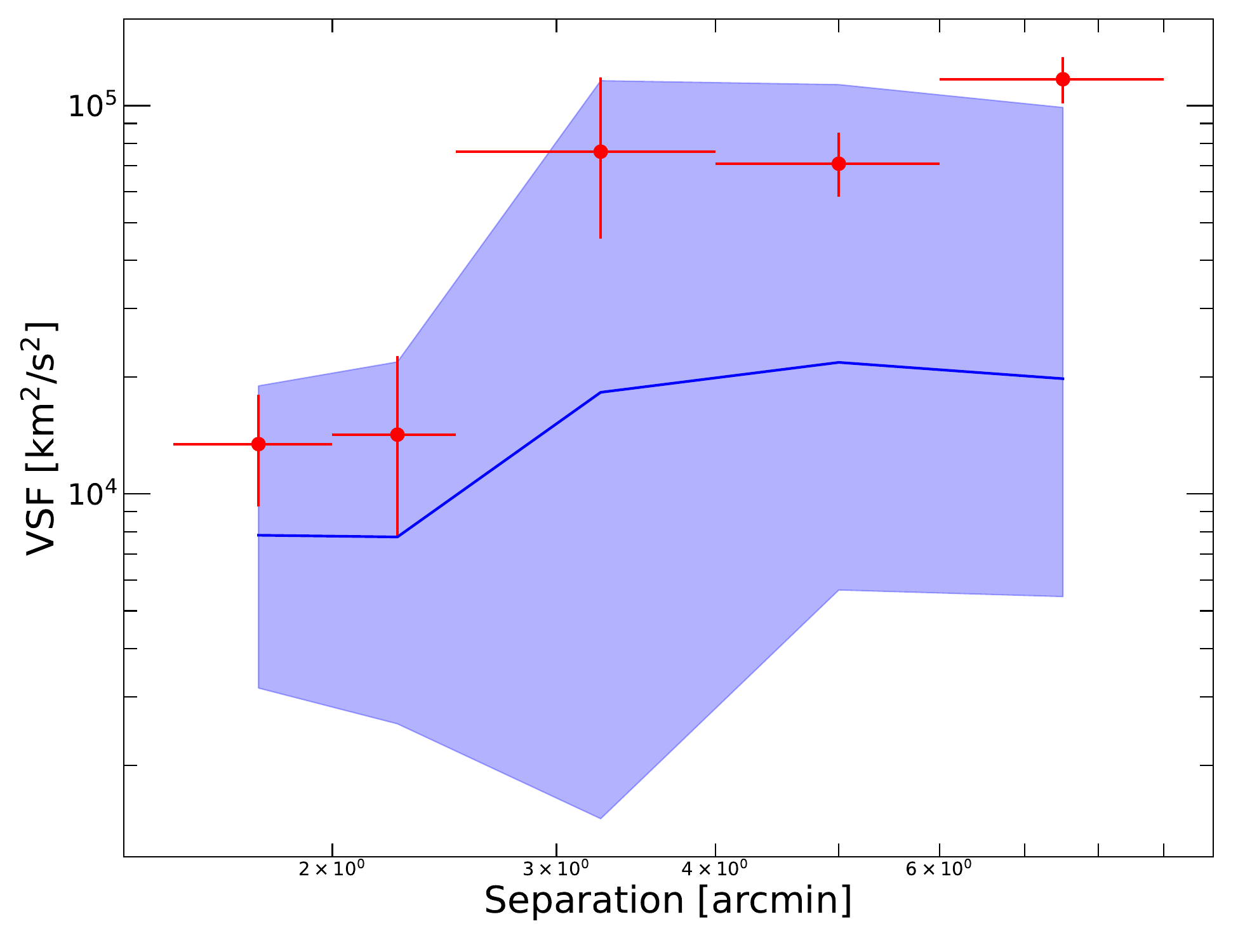}\includegraphics{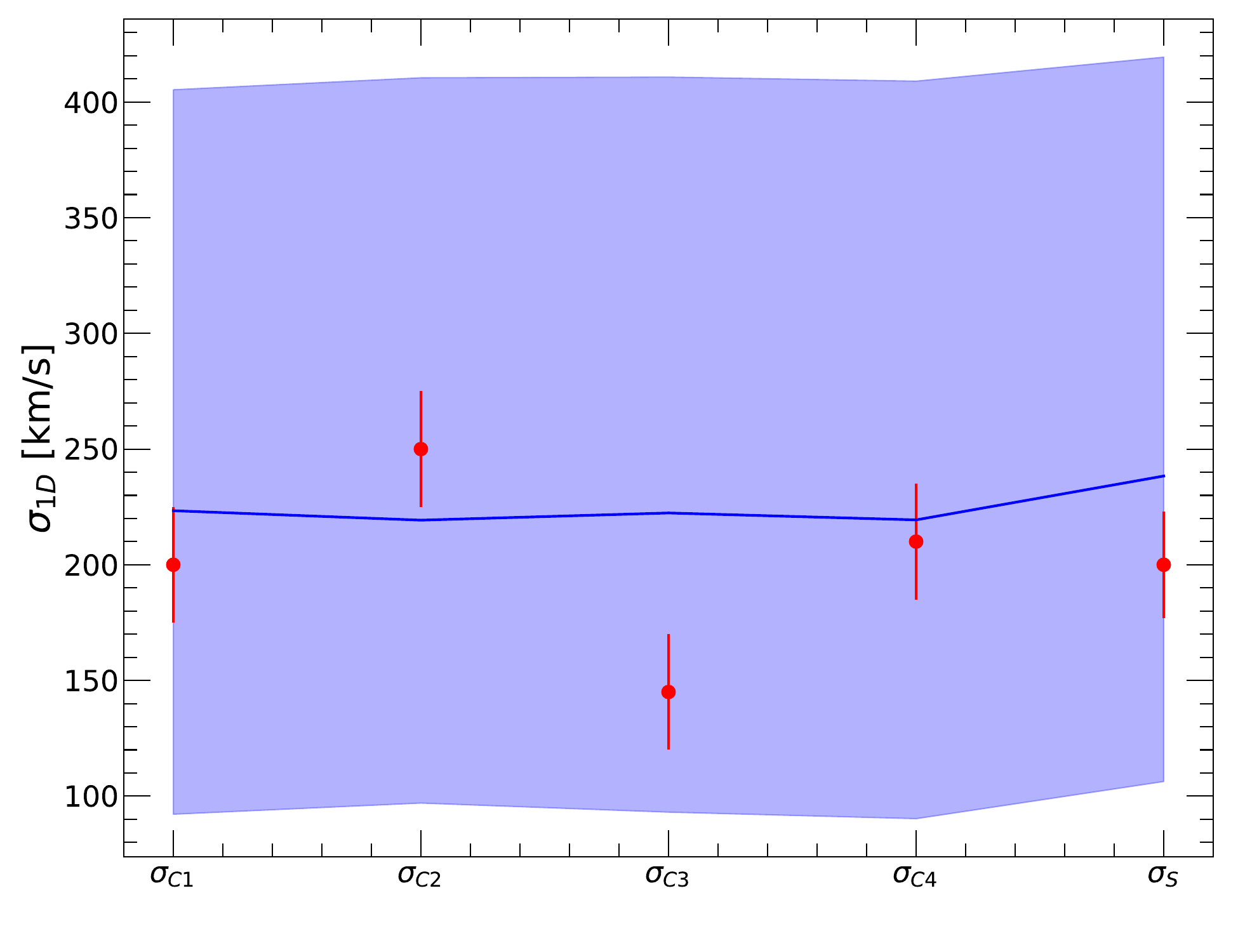}\includegraphics{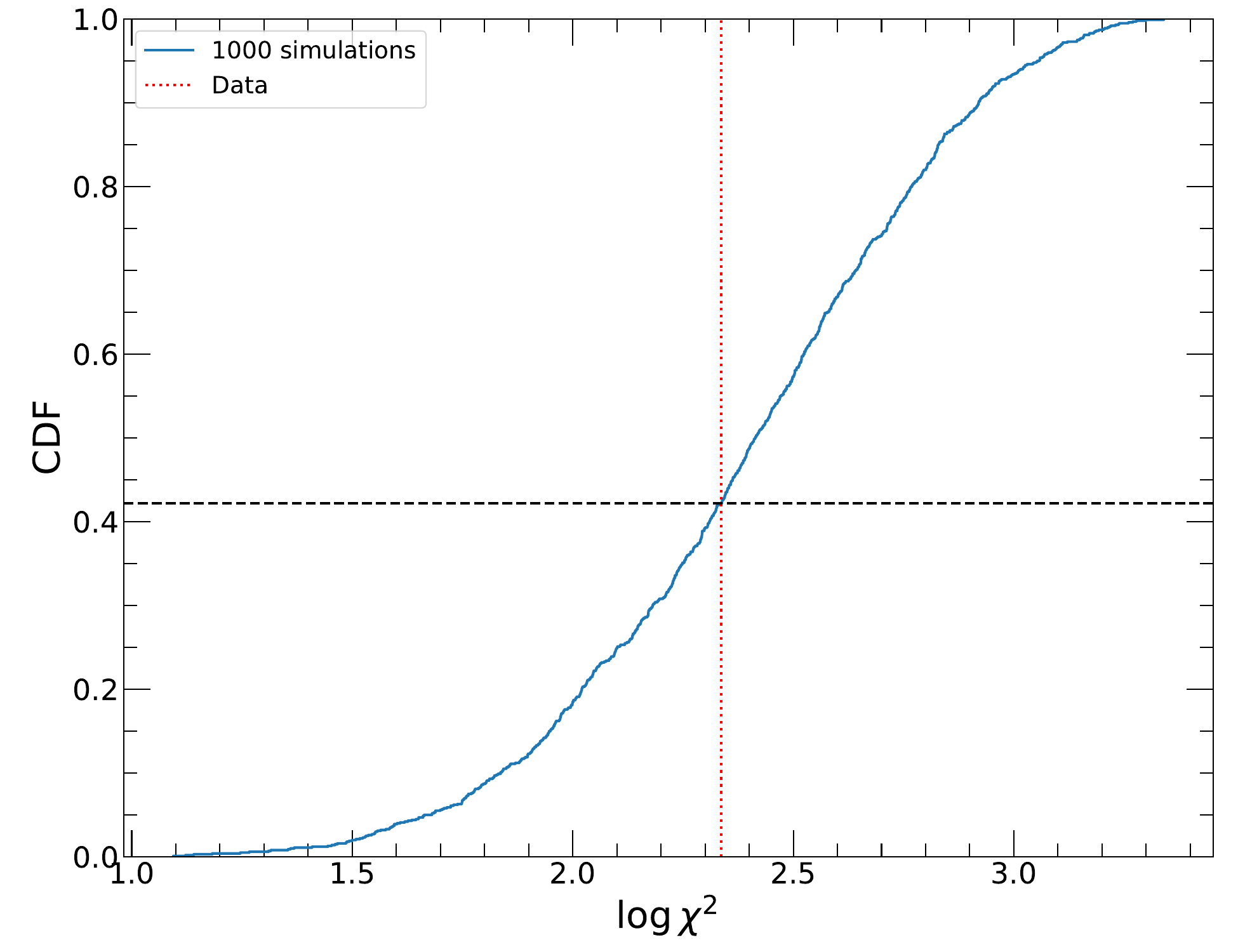}}
\caption{\label{fig:goodness}Goodness-of-fit tests applied to the Coma cluster \emph{XRISM}/Resolve data. The left and middle panels show the comparison between the Resolve data (red data points) and the best fitting model for the VSF (left) and the velocity dispersion measurements in the four quadrants of the central pointings and the full southern pointing (middle). The blue curves represent the median of 1,000 simulations generated from the posterior distribution, whereas the blue shaded areas show the 16th to 84th percentiles of the generated mock datasets. The right-hand panel shows the cumulative distribution of test statistic value (Eq. \ref{eq:teststat}) for 1,000 simulations generated from the posterior, with the test statistic value obtained for the data indicated as the dotted vertical line. The dashed black line indicates the fraction of simulations with a better test statistic value than the one obtained for the data.}
\end{figure*}

To check whether the model provides an appropriate description of the data, we extracted 1,000 random parameter sets from the posterior distribution of our fiducial three-parameter model (see Fig. \ref{fig:posterior}) and generated mock data vectors for each parameter set. We then studied the distribution of the mock datasets and attempted to compare the simulated data with the observations. However, as discussed in Sect. \ref{sec:optim} the individual bulk velocities fluctuate considerably from one simulation to another, as the simulated random velocity fields can generate both positive and negative fluctuations, which makes it difficult to directly compare the distribution of simulated bulk velocities with the measured values. To alleviate this issue, we computed the VSF (Eq. \ref{eq:vsf}) from each simulation and compared the distribution of mock VSF values to the observed VSF. 

In the left-hand panel of Fig. \ref{fig:goodness} we show the VSF obtained by combining the eight individual bulk velocity measurements shown in Fig. \ref{fig:coma_map} and compare the result with the median and dispersion of the mock VSF generated from the posterior parameter sets. We see that the median of the mock datasets lies below the observed values. However, the generated VSF values span a broad range of values which encompass the measured VSF values within $1\sigma$. As noted in X25, this is due to cosmic variance, as the existing measurements only cover a small fraction of the cluster's volume and the velocity field substantially fluctuates from one region to another. In the middle panel of Fig. \ref{fig:goodness} we show a similar comparison for the velocity dispersions averaged over the five regions considered (four quadrants of the central pointing and the full southern pointing). Here we can see that the data points lie very close to the median of the 1,000 models. The model predicts that the velocity dispersions could vary in the range 100-400 km/s for similar power spectrum parameters. Therefore, if the power spectrum parameters retrieved here accurately represent the true power spectrum, the measured velocity dispersions correspond to the typical expected values, whereas the bulk velocities fluctuate more than the typical value, although such fluctuations are not unlikely given the large cosmic variance. 

To test whether the observed deviation of the VSF from the median value matches the expectations of the model, following \citet{vwk25} we attempted to build a test statistic that quantifies the expected level of deviations between the model and the data. Since the uncertainties in the fitted velocities are approximately Gaussian, a natural choice for the test statistic is the $\chi^2$ distribution. However, in the case of the VSF we found that the distribution of generated values follows a log-normal distribution rather than a Gaussian distribution, such that we build a $\chi^2$ distribution in log space instead of real space. For each of the data points in Fig. \ref{fig:goodness} (left and middle panels), we calculated the mean value of the mock datasets, $\langle\sigma_{1D}\rangle$ and $\langle\ln\rm{VSF}\rangle$. We then defined our test statistic for each realization as
\begin{equation}
\chi^2 = \sum_{i\in \sigma_{1D}}  \frac{(\sigma_{1D,i} - \langle\sigma_{1D,i}\rangle)^2}{\Delta\sigma_{1D,i}^2} + \sum_{i\in{\rm VSF}} \frac{(\ln\rm{VSF}_i - \langle\ln\rm{VSF}_i\rangle)^2}{\Delta\ln\rm{VSF}_i^2},
\label{eq:teststat}
\end{equation}
where the summation is done over the velocity dispersion ($\sigma_{1D}$) and VSF points, and the denominators represent the statistical uncertainties in the velocity dispersion and VSF, respectively. 

We computed the value of our test statistic for the 1,000 individual realizations of the posterior distribution and built the cumulative distribution of the $\chi^2$ values. The resulting distribution is shown in the right-hand panel of Fig. \ref{fig:goodness}. We performed the same calculation for the observed dataset and compared the retrieved test statistic value with the distribution of the 1,000 random datasets. We found that the measured $\chi^2$ value is lower than the mock $\chi^2$ value for 58\% of the datasets, such that the observed deviations are fully consistent with the expectations. Thus, we conclude that the model provides an adequate description of the data at hand. Obviously, this conclusion only holds for the currently available dataset, and additional constraints providing a wider spatial coverage or a higher spatial resolution may still require a more complex model.

\section{Discussion}
\label{sec:disc}

\subsection{Power spectrum reconstruction method}

The recovery tests shown in Sect. \ref{sec:recovery} demonstrate that the method proposed here allows to retrieve interesting constraints on the velocity fluctuation power spectrum in galaxy clusters, even in the presence of a coarse observational sampling. Our results show that in the case of the Coma cluster, two XRISM/\emph{Resolve} pointings are sufficient to get a reliable estimate of the turbulent Mach number and interesting constraints on the injection scale, although a wider coverage is required to constrain the slope of the turbulent cascade. Compared to surface brightness fluctuation techniques \citep[e.g.][]{zhuravleva14,dupourque23}, our study directly traces the underlying velocity field without requiring assumptions on the link between surface brightness and velocity fluctuations. The use of simulations allows us to include a number of important observational effects that are difficult to implement analytically, such as projection effects, the emissivity distribution of the source, and the PSF of the instrument. Our code also allows us to implement the exact spatial footprint of the available observational datasets.

We stress that our method naturally takes cosmic variance into account, since the large training sample includes fluctuations in the local measurements induced by the varying large-scale fluctuation fields. At this point, cosmic variance is the largest source of uncertainty in the reconstruction of the velocity power spectrum because of the limited spatial coverage. Additional \emph{XRISM}/Resolve pointings are needed to constrain the velocity field over a wider spatial range and decrease the cosmic variance, which is the current limiting factor for the reconstruction of the turbulent Mach number and the injection scale (see Fig. \ref{fig:goodness}). For instance, \citet{zuhone16} originally proposed a five-pointing strategy with one pointing in the center and four equally spaced offset pointings located 200 kpc from one another. This strategy greatly reduces cosmic variance with respect to the current configuration and probes a wider range of scales, from 40 to 400 kpc, which allows us to break the degeneracy between injection scale and slope. Three additional pointings would be sufficient to distinguish unambiguously between a small (few hundred kpc) and a large ($\gtrsim1$ Mpc) injection scale. 

Our method naturally takes into account a potential offset between the local gas bulk velocities and the rest frame of the system set by the mean redshift of the member galaxies. This is crucial in the case of Coma, where large bulk velocities between the gas and the galaxies (up to $\sim800$ km/s) are observed. Conversely, approaches relying exclusively on the velocity structure function \citep[e.g.][]{zuhone16,molin25} require a wider spatial coverage of the system to provide unbiased estimates of power spectrum parameters, as the VSF only depends on relative bulk velocities rather than the absolute ones. Correspondingly, our recovery tests show that with a limited number of available measurements the VSF only provides lower limits to the true turbulent Mach number. In X25, we circumvented this problem by including an additional VSF point to represent the velocity difference between the gas and the galaxies, although the scale had to be fixed at an arbitrary value of 1 Mpc. Future instruments such as the X-ray Integral Field Unit \citep[X-IFU,][]{xifu} on board \emph{NewAthena} \citep{newathena} will allow us to measure gas velocities at higher spatial resolution and over wider areas, which will provide constraints on the Mach number independently of any external measurement. These measurements will also probe a much wider range of spatial scales, which will allow us to determine the slope of the turbulent cascade and potentially detect the dissipation scale. 

\subsection{Model power spectrum}

Applying our technique to the set of two \emph{XRISM}/Resolve pointings of the Coma cluster presented in X25 (see Sect. \ref{sec:inference}), we found that the data can be well described by an isotropic turbulent cascade model with a 3D Mach number of $\sim0.7$ and a large injection scale ($>1$ Mpc). Considering the expected level of cosmic variance in a two-pointing configuration, the observed data are consistent with the expected level of fluctuations around the mean value (Fig. \ref{fig:goodness}), such that the model provides an adequate description of the data at hand. In Table \ref{tab:res} we provide the best fitting parameters for the fiducial 3-parameter model as well as for several other configurations (see below). Here we discuss the implications of these results in case the underlying assumptions of a single clump with an isotropic velocity field are born out.

\begin{figure}
\resizebox{\hsize}{!}{\includegraphics{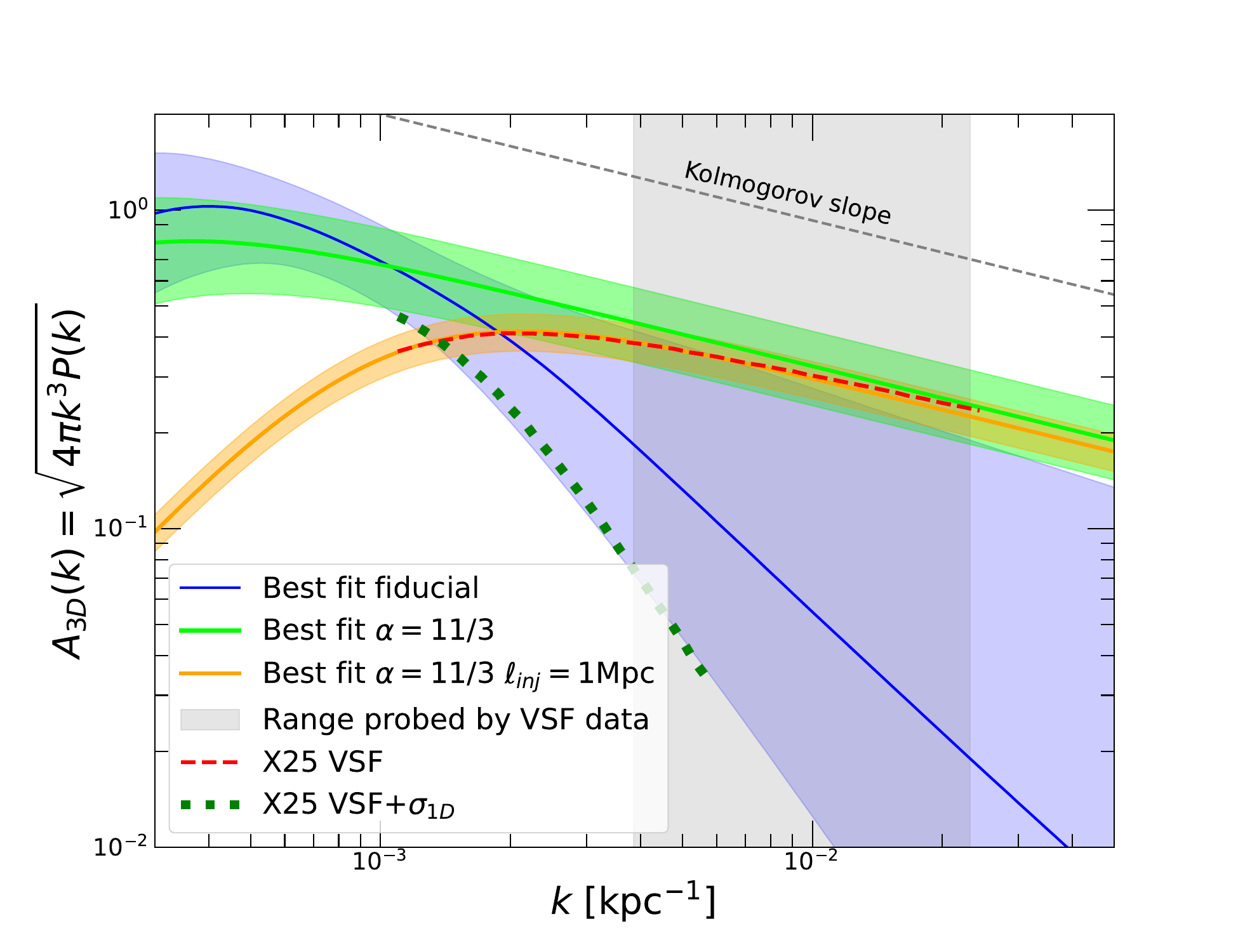}}
\caption{\label{fig:fitted_ps}Amplitude of velocity fluctuations as a function of wave number for the Coma cluster. The curves are calculated from the posterior distribution of model parameters for a model with free power-law slope (blue), a model with Kolmogorov slope $\alpha=11/3$ (light green), and a model with Kolmogorov slope and $\ell_{inj}=1$ Mpc (orange). The solid curves and the shaded areas show the median and the $1\sigma$ error envelope of the fitted models. The dashed red curve shows the X25 model fitted to the VSF only, whereas the dotted green curve represents the fit to the VSF and velocity dispersion data from X25. For comparison, the gray shaded area shows the range of scales probed by the bulk velocity data, whereas the gray dashed line indicates the Kolmogorov slope.}
\end{figure}

In Fig. \ref{fig:fitted_ps} we show the posterior envelope of the model power spectra converted into fluctuation amplitudes \citep{churazov12},
\begin{equation}
A_{3D}(k) = \frac{v_{3D}(k)}{c_s} = \sqrt{4\pi k^3 P_{3D}(k)}
\end{equation}
with $P_{3D}$ defined as a function of the 3D Mach number rather than the velocity dispersion (see Fig. \ref{fig:posterior}). As discussed in X25, the large bulk velocities observed in Coma exceed the values measured in other clusters observed by \emph{XRISM} (Perseus, A2029, Centaurus), where the bulk gas velocities reach a maximum of 300 km/s with respect to the mean of the galaxy redshifts \citep{xrismcentaurus,xrisma2029b}. Our modeling implies that Coma is much more dynamically active than the aforementioned clusters, which are all classified as morphologically relaxed systems with a central cool core \citep{campitiello22}. 

In Fig. \ref{fig:fitted_ps} we compare our best-fitting models with the results presented in X25, where the VSF and the velocity dispersions were calculated directly from the inverse Fourier transform of the model power spectrum. The injection scale was fixed to 1 Mpc and an additional point at an arbitrary scale of 1 Mpc was added to the VSF to represent the velocity difference between the gas and the galaxies. Our results are statistically consistent with the X25 results, which lie within the 1$\sigma$ range. To allow for a direct comparison between this method and the results of X25, we ran an additional model reconstruction with $\ell_{inj}$ fixed to 1 Mpc and Kolmogorov slope, leaving only the normalization as a free parameter. The retrieved power spectrum is in excellent agreement with the inference of the same model from X25. However, the goodness-of-fit test (Sect. \ref{sec:goodness}) shows that the single-parameter provides a substantially worse description of the data, which motivates keeping the injection scale or the power-law slope as a free parameter. We also considered the model power spectrum for a configuration where the injection scale is fixed to 1 Mpc and the slope is left free to vary. In this case, we retrieve a very steep slope ($\alpha=7.45_{-1.25}^{+0.46}$), which hits the boundary of the prior, in agreement with the discussion in X25. 

\subsection{Non-thermal pressure support}

\begin{table*}
\caption{\label{tab:res}Best fitting power spectrum parameters for the Coma cluster for various models.}
\begin{center}
\begin{tabular}{lccccc}
\hline
Model &  $\ell_{inj}$ [Mpc] & $\alpha$ & $\mathcal{M}_{3D,\rm tot}$ & $\mathcal{M}_{3D}(>k_{500})$ & $\mathcal{M}_{3D}(>k_{200})$ \\
\hline
\hline
Fiducial & $2.2_{-1.0}^{+2.0}$ & $5.3_{-1.0}^{+2.1}$ & $0.72_{-0.22}^{+0.28}$ & $0.45_{-0.13}^{+0.18}$ & $0.57_{-0.14}^{+0.17}$\\
Fixed $\alpha$ & $6.6_{-2.2}^{+2.5}$ & 11/3 & $0.73_{-0.20}^{+0.25}$ & $0.60_{-0.16}^{+0.18}$ & $0.66_{-0.18}^{+0.20}$\\
Fixed $\ell_{inj}$ & 1 & $7.45_{-1.25}^{+0.46}$ & $0.55_{-0.14}^{+0.22}$ & $0.37_{-0.09}^{+0.11}$ & $0.45_{-0.10}^{+0.14}$\\
Fixed $\alpha$, $\ell_{inj}$ & 1 & 11/3 &  $0.45_{-0.06}^{+0.06}$ & $0.44_{-0.05}^{+0.05}$ &  $0.45_{-0.06}^{+0.06}$\\
\end{tabular}
\end{center}
\tablefoot{The total Mach numbers are estimated as the integral of the power spectrum (Eq. \ref{eq:classicalps}) whereas the values estimated within various ranges of scales are computed over all wavenumbers greater than a limiting wavenumber $k_{500}=1/R_{500}$ and $k_{200}=1/R_{200}$.}
\end{table*}

In case the velocity field is fully isotropic, the turbulent Mach number can be linked to the non-thermal pressure fraction in the system as \citep{eckert19}
\begin{equation}
\frac{P_{NT}}{P_{\rm tot}} = \frac{\mathcal{M}_{3D}^2}{\mathcal{M}_{3D}^2 + \frac{3}{\gamma}},
\label{eq:pnt}
\end{equation}
\noindent with $\gamma=5/3$ the adiabatic index. It is important to note that, given the large injection scale and/or steep turbulent cascade, most of the power in our best fitting model lies in the very large scales ($\ell\gtrsim2$ Mpc), which rival or even exceed the virial radius of the system. These large-scale motions are not directly probed by our model and, if present, they do not contribute to the kinetic pressure within the halo. We recall that the turbulent Mach number inferred in our model is defined as the integral of the power spectrum (Eq. \ref{eq:classicalps}) over all scales, which includes a significant contribution from scales larger than 2 Mpc that are not directly probed in this study and may not even contain much gas. Therefore, it is worthwhile estimating the Mach number only within the scales that are relevant to the pressure balance within $R_{500}$ and/or $R_{200}$. From the mass-temperature relation of \citet{umetsu20} and a temperature of 8.37 keV (X25), we retrieve $R_{500}=1400$ kpc and $R_{200}=2153$ kpc. These estimates agree within 5\% with the values obtained from the \emph{Planck} $Y_{SZ}$ measurement \citep{planck13_coma}. The effective turbulent Mach number within $R_{500}$ can be estimated as the integral of the power spectrum over all wavenumbers greater than $k_{500}=1/R_{500}$, i.e.
\begin{eqnarray}
\mathcal{M}_{3D,500} & = & \left[ \int_{k_{500}}^\infty 4\pi k^2 P_{3D}(k)~{\rm d}k\right]^{1/2}\\
\, & = & \mathcal{M}_{3D,\rm tot}  \left[  \frac{\int_{k_{500}}^\infty k^2 (1+(k/k_{inj})^2)^{-\alpha/2}~{\rm d}k}{\int_{0}^\infty k^2 (1+(k/k_{inj})^2)^{-\alpha/2}~{\rm d}k} \right]^{1/2} .
\end{eqnarray}
For our fiducial model, the Mach number integrated within $R_{500}$ decreases to $\mathcal{M}_{3D,500}=0.45_{-0.13}^{+0.18}$. The non-thermal pressure contribution within $R_{500}$ can be estimated from the posterior chains on the power spectrum parameters by calculating the range of values of $\mathcal{M}_{3D,500}$ and inferring the 16th and 84th percentiles of the non-thermal pressure ratio distribution (Eq. \ref{eq:pnt}), which yields
\begin{equation}
\frac{P_{NT}}{P_{\rm tot}}(<R_{500}) = 0.10_{-0.04}^{+0.08}.
\end{equation}
This estimate considers only the scales that effectively contribute to balancing gravity within the halo. In Table \ref{tab:res} we report the best-fit parameters for the three models considered here (fiducial, Kolmogorov slope, fixed injection scale) as well as the turbulent Mach numbers estimated within $R_{500}$ and $R_{200}$. Within $R_{500}$, all the models retrieve a Mach number in the range $0.4-0.6$, which exceeds the value estimated directly from the broadening of the emission lines by a factor of $\sim2$ ($\mathcal{M}_{3D}=0.24$ from small-scale motions in X25). Jointly modeling the velocity dispersions and the bulk velocities allows us to probe the full range of spatial scales, which leads to a larger kinetic energy fraction compared to the values retrieved from the line-of-sight velocity dispersion. Indeed, the estimated 1D velocity dispersion is weighted by the local gas emissivity (Eq. \ref{eq:sigmav1D}), which varies substantially along the line of sight. Even in the case of the Coma cluster, which features a flat emissivity distribution within its core, 90\% of the emission originates from the innermost 600 kpc, such that large-scale modes do not contribute to the measured line broadening. 

We stress, however, that the concept of non-thermal pressure is only meaningful in case the cluster is essentially virialized and the acceleration term in the Euler equation is negligible \citep[e.g.][]{lau09,vazza18,angelinelli20}, which may not hold in the early stages of a major merger.

\subsection{Limitations of the model}

In the standard turbulent cascade scenario, the injection scale is expected to be comparable to the size of the main perturbing body. The very large injection scale inferred by our model is comparable to the size of the cluster itself, which makes it unrealistic in cosmological structure formation scenarios \citep[e.g.][]{vazza12b,va17turb}. The large injection scale inferred for the Coma cluster may represent a velocity field that includes a major merger between two cluster-scale entities with characteristic sizes greater than a Mpc. This picture is consistent with the simulations of \citet{vazza12b}, who showed that velocity power spectra in clusters do not show a clear cut-off at low wave numbers (and can be formally described by a high injection scale) and are dominated by ongoing mergers on large scales and turbulence on smaller scales ($\ell\lesssim0.3R_{200}$). Coma is known to be a dynamically active system, as it features a pair of concentric shock fronts located $\sim1$ Mpc East and West of the cluster core \citep{planck13_coma,churazov21} and a giant radio halo \citep{brown11}. The large bulk velocities measured in the two Resolve pointings imply that there is a substantial line-of-sight component to the velocity field, although the presence of the concentric shock fronts requires plane-of-the-sky motions as well. Alternatively, if the merger occurs preferentially on the plane of the sky, the gas may be compressed along the merger axis and form perpendicular outflowing gas ``tongues'', as observed for instance in Abell 754 \citep{botteon24}. Comparing the observed Coma velocity field and thermodynamic properties with a grid of merger simulations with different mass ratios and impact parameters is necessary to test this scenario \citep[e.g.][]{zuhone18}. 

Obviously, our method relies on the assumption that the velocity field can be accurately described by a Gaussian random field. This assumption is not necessarily satisfied in case the merger is in its early phase, where the velocity difference between the two merging entities is the primary source of large-scale bulk motions. If so, we could be observing the system before the turbulent cascade is fully developed and the power spectrum may be temporarily much steeper than Kolmogorov (see the discussion in X25). In this case, a two-component model including the bulk motions on large scales and the previously generated turbulent cascade on smaller scales may be more physically motivated than the single-component isotropic model considered here. Additional pointings covering a wider area are required to test this scenario.

\section{Conclusion}

In this paper, we proposed a method to retrieve the shape of the velocity fluctuation power spectrum in galaxy clusters based on measurements of X-ray line shifts and line broadening. We then applied our model to the \emph{XRISM}/Resolve observations of the Coma cluster to study the shape of the velocity power spectrum in this system, estimate the non-thermal pressure fraction, and constrain the merger scenario. Our results can be summarized as follows:

\begin{itemize}

\item We showed that the velocity fluctuation power spectrum can be estimated by generating simulations of a Gaussian random field and forward modeling the measured bulk velocities and velocity dispersions. This approach allows us to include several important observational effects (e.g., projection effects, emissivity weighting, PSF smearing) and generate realistic mock datasets. 

\item Given a parametric form for the velocity power spectrum, our method generates a large number of simulations from random parameter sets and trains a neural network to learn the mapping between the model power spectrum parameters and the mock data vectors. This is achieved through the SNPE algorithm implemented in the \texttt{sbi} package \citep{sbi}. We distribute the code in the form of an easy-to-use Python package which can be applied to any observational configuration and emissivity distribution.

\item We considered a two-pointing \emph{XRISM}/Resolve configuration tailored to the available observations of the Coma cluster (X25). We tested the ability of our code to recover the input power spectrum parameters for a simple paramterization of the power spectrum considering a turbulent velocity field characterized by an injection scale and a turbulent cascade towards smaller scales. We found that two pointings are sufficient to accurately determine the turbulent Mach number and provide interesting estimates of the injection scale. However, a finer sampling is needed to constrain the slope of the turbulent cascade.

\item We compared two different configurations for the provided data vectors. In the first case, we fit directly for the individual bulk velocities, whereas in the other case we compute the VSF from each generated simulation and compare the results with the measured VSF. We showed that the former configuration provides tighter constraints on the turbulent Mach number, as the absolute bulk velocities include additional information on the motion of the gas with respect to the cluster's rest frame.

\item Applying our method to the Resolve measurements, we found that a power spectrum model with $\mathcal{M}_{3D,\rm tot}=0.72_{-0.22}^{+0.28}$ and $\ell_{inj}=2.2_{-1.0}^{+2.0}$ Mpc provides a good description of the available data. Specifically, a relatively high Mach number together with a large injection scale are required to reproduce the large bulk velocities but low velocity dispersions observed in the two Resolve pointings. Such large injection scales are unexpected in the usual cosmological structure formation, which implies that the system is likely in a merging phase with a substantial line-of-sight component.

\item Our power spectrum model is in full agreement with the one derived in X25 from the same data but using a different method, based on fitting the velocity structure function. In comparison, our method allows for a natural inclusion of the external information on galaxy velocities. 

\item The inferred turbulent Mach number within $R_{500}$ estimated with our model ($\mathcal{M}_{3D,500} = 0.45_{-0.13}^{+0.18}$) is noticeably larger than the value estimated from the line broadening alone ($\sqrt{3}\mathcal{M}_z=0.24-0.25$). The difference can be explained by the joint modeling of line shifts and line broadening, which given the large bulk velocities requires a large injection scale and/or a steep turbulent cascade. Additional pointings covering a wider area are required to reduce the cosmic variance and confirm our results.
\end{itemize}

\begin{acknowledgements} 
DE and MR acknowledge support from the Swiss National Science Foundation (SNSF) under grant agreement 200021\_212576. Support for JAZ was provided by the Chandra X-ray Observatory Center, which is operated by the Smithsonian Astrophysical Observatory for and on behalf of NASA under contract NAS8-03060. IZ acknowledges partial support from the Alfred P. Sloan Foundation through the Sloan Research Fellowship and from NASA grant 80NSSC18K1684. NT acknowledges the support by NASA under award number 80GSFC24M0006. NO acknowledges JSPS KAKENHI Grant Number JP25K07368. DRW acknowledges support from NASA grant 80NSSC23K0740.
\end{acknowledgements}

\bibliographystyle{aa}
\bibliography{clusterps}

\appendix

\section{Surface brightness distribution}
\label{sec:sb}

\begin{figure}
\resizebox{\hsize}{!}{\includegraphics{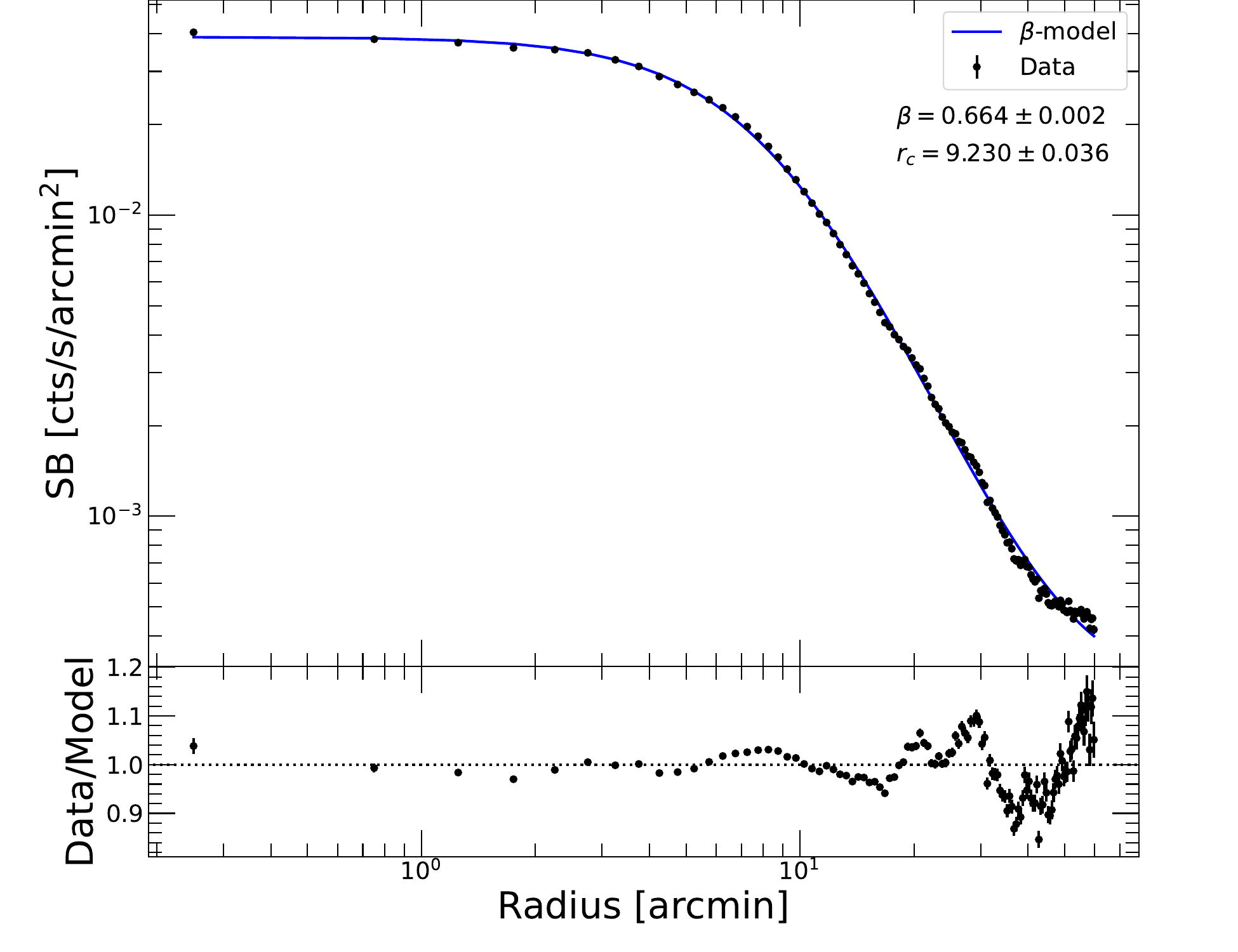}}
\caption{\label{fig:sb}\emph{ROSAT}/PSPC surface brightness profiles of the Coma cluster (black data points) modeled with a spherical $\beta$-model (Eq. \ref{eq:betamodel}, blue curve). The bottom panel shows the residuals quantified as the ratio of data to model.}
\end{figure}

The results presented here make important assumptions on the true underlying emissivity distribution. That is, we assume that the emissivity distribution is spherically symmetric and can be described by a single $\beta$-model (Eq. \ref{eq:betamodel}). This assumption is supported by the surface brightness distribution of the gas in Coma, which appears relatively symmetric out to $\sim30$ arcmin with a large central core. In Fig. \ref{fig:sb} we show the surface brightness profile of Coma extracted from \emph{ROSAT}/PSPC data because of the instrument's large FoV \citep{briel92}. The profile was extracted from a mosaic made of three pointed PSPC observations covering the cluster out to $\sim1.5$ degrees ($\sim2.5$ Mpc) using the Python package \texttt{pyproffit} \citep{eckert20}. While the $\beta$-model is too simplistic to reproduce all the features observed in the data, it provides a good description of the brightness distribution, with deviations at the level of $\leq10\%$ out to $\sim50$ arcmin (see the bottom panel of Fig. \ref{fig:sb}). In the regions where the \emph{XRISM}/Resolve observations are obtained, the model predicts that 90\% of the emission originates from a region of $\sim600$ kpc length. 

While the spherical $\beta$-model provides an adequate description of the 1D profile, the isophotes are elongated along the E-W axis, such that an elipsoidal distribution likely provides a better description of the 3D cluster shape. To study the impact of ellipticity on the results presented in this work, we extracted the brightness profile in elliptical annuli, with the shape of the elliptical annuli matching the ellipticity ratio of $\sim1.2$ between the major and minor projected axes observed in the X-ray isophotes. We then assumed that the third (line-of-sight) dimension has a similar shape as the projected major axis. We then used the corresponding 3D distribution to generate a new set of cluster simulations and train another model. With this configuration, we obtain $\mathcal{M}_{3D}=0.80_{-0.24}^{+0.20}$, $\alpha=6.0_{-1.7}^{+1.4}$, and $\log k_{inj}=-3.34_{-0.34}^{+0.19}$, which is consistent with the results obtained from a spherical emissivity distribution within $1\sigma$.

\end{document}